# FREQUENCY AND VIABILITY OF DIPLOID AND HAPLOID MALES OFFSPRINGS OF MATED FEMALES OF SOLITARY ENDOPARASITOID (*DIADROMUS PULCHELLUS* ) ( ICHNEUMONIDAE) .

**Robert Kalmès \*, Danielle Rojas-Rousse**

Institut de recherches sur la biologie de l'insecte –UPRES A 6035, faculté des sciences et techniques avenue Monge - 37200 Tours, France

Correspondenceand reprints.
E-mail address :: kalmes@univ-tours.fr (R. Kalmès)

**Abstract** - Sex determination in the order Hymenoptera is based on arrhenotoky, hymenopteran males are usually haploid and females diploid. Males of the Ichneumonidae *Diadromus pulchellus* , solitary endoparasitoid of *A. assectella* pupae are normally haploid, but diploid males are present in a natural population and can be obtained in a experimental population. The future of an ovocyte laid by mated females of the solitary endoparasitoid *Diadromus pulchellus* was characterised by 6 probabilities related to the sex and the development of the ovocyte. The probabilities of fertilisation of female ovocyte ($k_1$) or non-fertilisation ($k_3$) showed that an inseminated female functioned as a unmated female for half of the time (since $k_1$= 0.492 and $k_3$ = 0.455) with the probability of fertilisation of male ovocyte _($k_2$ = 0.053).The survival probabilities of each type of ovocyte showed that an ovocyte had a high probability of developing up to the adult stage, although the difference between the calculated sex ratios at laying  (males / females = 1.032) and at emergence (0.90) revealed a slight reduction in the number of haploid sons.

The probabilities of fertilisation and of viability of all the ovocytes laid by each of the 33 mated females were analysed by an ascending hierarchical classification of Euclidean distances and by an analysis of their principal components. The 33 mothers were distributed into 4 distinct sub-groups characterised by a sex ratio varying from an exclusive presence of females to an exclusive presence of males. Our hypothesis was that this distribution in 4 sub-sets could not simply result from the random nature of the sample.

**solitary endoparasitoid / probabilities of fertilisation / survival probabilities / haploid male / diploid male / sex-ratio.**

**Résumé - Fréquences et viabilité des mâles diploïdes et haploïdes de femelles inséminées de l'endoparasitoïde solitaire (*Diadromus pulchellus*) ( Ichneumonidae).** Le mode de reproduction fondamental des Hyménoptères est basé sur la parthénogenèse arrhénotoque, donnant des mâles haploïdes, tandis que les femelles diploïdes résultent de la fécondation d'un ovocyte Chez *Diadromus pulchellus*, un endoparasitoïde solitaire de la chrysalide de la teigne du poireau (<u>Acrolepiosis assectella</u>) l'étude de populations naturelles et expérimentales a révélé la présence de mâles diploïdes. Le devenir d'un ovocyte pondu par des femelles accouplées de *Diadromus pulchellus* est cacartérisé par 6 probabilités liées au sexe et au développement d'un ovocyte. Les probabilités de fertilisation ($k_1$) et de non fertilisation ($k_3$) indiquent qu'une femelle inséminée fonctionne pour moitié comme une femelle non accouplée car ces deux probabilités sont proches ($k_1 = 0.492$ et $k_3 = 0.455$). Une femelle fertilise les ovocytes de sexe mâle avec une probabilité ( $k_2 = 0.053$ ). Les estimations attachées à la survie montrent qu'un ovocyte pondu à de fortes probabilités d'évoluer jusqu'à l'adulte bien que l'écart entre le sex-ratio à la ponte (mâles / femelles = 1.032) et celui à l'émergence (0.90), révèle un déficit en mâles haploïdes. Un classement hiérarchique des 6 estimations répartit les femelles dans 4 familles caractérisées par la variabilité du sex-ratio allant d'une présence exclusive de filles à celle exclusive de fils. Cette répartition des femelles dans 4 ensembles ne permet pas d'exclure l'hypothèse par laquelle la classification résulterait uniquement de la structure aléatoire de l'échantillon de femelles testées.

**endoparasitoïde solitaire / probabilité de fertilisation / probabilité de survie /mâle**

**haploïde / mâle diploïde / sex-ratio.**

**Version abrégée**

Le but de cette note est de déterminer la fréquence de fertilisation ou non d'un ovocyte pondu par les femelles accouplées de *Diadromus pulchellus*, endoparasitoide solitaire de la chrysalide de la teigne du poireau (*Acrolepiopsis assectella*),et sa viabilité. Ces estimations ont porté sur toute la ponte comprise de la naissance au pic reproducteur des femelles et, ont nécessité un modèle de probabilités tenant compte de la fertilisation ou non d'un ovocyte et de sa viabilité qu'il soit de sexe femelle ou mâle diploïde ou mâle haploïde.

Un premier ensemble de trois probabilités ($k_1$ - $k_2$ - $k_3$) est attaché au sexe de l'ovocyte pondu et, un deuxième ensemble ($w_1$ - $w_2$ - $w_3$) est affecté à la viabilité des ovocytes pondus , de sexe déterminé

**$k_1$** est la probabilité qu'un ovocyte soit fécondé par un spermatozoïde et donne un oeuf de sexe femelle ;

**$k_2$** est la probabilité qu'un ovocyte soit fécondé par un spermatozoïde et donne naissance à un oeuf de sexe mâle. A priori $k_1$ et $k_2$ sont supposés différents .

**$k_3$** est la probabilité complémentaire ($k_3 = 1 - k_1 - k_2$) qu'un ovocyte pondu par une femelle accouplée ne soit pas fécondé et s'engage dans un développement parthénogénétique :il sera haploïde et de sexe mâle.

**$w_1$** est la probabilité qu'un oeuf diploïde femelle donne un adulte vivant femelle diploïde ;

**$w_2$** est la probabilité qu'un oeuf diploïde mâle donne un adulte vivant mâle diploïde ;

**$w_3$** est la probabilité qu'un oeuf haploïde mâle donne un adulte vivant mâle haploïde que la femelle mère soit accouplée ou vierge. C'est la probabilité la plus accessible. Elle est évaluée à partir de la ponte des femelles vierges où tous les ovocytes, pondus sont non fécondés et ont la potentialité de donner des mâles adultes haploïdes. Cette probabilité $w_3$, calculée grâce à la ponte des femelles vierges, est aussi affectée aux œufs haploïdes que peut pondre une femelle accouplée. Il s'agit donc d'une valeur constante commune à toutes les femelles ( vierges et accouplées ).

Les descendances des 33 femelles accouplées sont caractérisées par la liste des probabilités $k_1$ - $k_2$ - $k_3$ et $w_1$ - $w_2$ - $w_3$. L'estimation de ces probabilités se fera sur le devenir des 962 ovocytes pondus par l'ensemble des femelles fécondées et pour chacune de ces femelles.

**I. Descendance des femelles accouplées: fréquence et viabilité des mâles diploïdes et haploïdes**

Sur l'ensemble femelles accouplées de *D. pulchellus*, l'estimation du paramètre de fertilisation [probabilités de fertilisation des ovocytes de sexe femelle et mâle ($k_1 + k_2 = 0,545$)] démontre que la moitié des ovoytes pondus est fertilisée. La valeur calculée de chaque paramètre révèle que la probabilité de la fertilisation d'un ovocyte de sexe mâle ($k_2 = 0,052$) est 10 fois plus faible que celle d'un ovocyte de sexe femelle ($k_1 = 0,492$).

A cette faible proportion d'ovocytes diploïdes de sexe mâle, s'ajoute celle des ovocytes non fertilisés ($k_3 = 0,456$) engagés dans la voie d'un développement parthénogénétique. Evaluées en période d'activité maximale de ponte, ces estimations montrent que les femelles accouplées de *D.pulchellus* fonctionnent pour moitié comme des femelles vierges : ovocytes de sexe femelle $k_1 = 0,492$ ; ovocytes de sexe mâle haploïde $k_3 = 0,456$. Ceci révèle qu'une femelle accouplée de parasitoïde peut gérer le stock de spermatozoïdes accumulés après un accouplement. Une étude chiffrée de cette gestion par les femelles de l'Eupelmide *Eupelmus orientalis*, montre l'utilisation progressive des spermatozoïdes en fonction de l'activité de ponte et de l'âge des femelles mères .Or jusqu'ici, les mâles haploïdes étaient supposés être les descendants exclusifs des femelles vierges et des femelles accouplées ayant en partie épuisé leur stock de spermatozoïdes Cette économie dans la gestion entraîne la présence de quelques filles en fin de vie.

A la ponte, le sex-ratio estimé proche de 1, [fils (2n + n) / filles soit 1,032], montre l'égalité phénotypique entre les sexes. Celle-ci masque un coût reproductif inégal entre les deux catégories de fils car la production d'un fils diploïde nécessite la mobilisation des spermatozoïdes et de toutes les sécrétions favorables à la fécondation de l'ovocyte .La ponte d'un ovocyte par hôte excluant la mortalité due à la compétition entre larves néonates, le développement de l'oeuf à l'adulte peut être malgré tout perturbé au stade embryonnaire ou larvaire ou nymphal. En absence de superparasitisme, un ovocyte pondu quelque soit son sexe, a une forte probabilité d'évoluer jusqu'au stade adulte car, la probabilité de survie d'un oeuf de sexe femelle est de 0,971, celle d'un oeuf de sexe mâle 2n de 0,783 et, celle d'un ovocyte non fécondé de 0,842. Le sex-ratio à l'émergence de 0,90 témoigne du faible niveau de mortalité puisqu'il prend en compte dans son calcul la probabilité de survie des ovocytes pondus. Ce sex-ratio à l'émergence révèle un faible déficit en adultes mâles haploïdes et diploïdes. Cela confirme l'observation que chez *D. pulchellus* même en absence de compétition larvaire la mortalité affecte en priorité les mâles

**2. Répartition dans quatre sous ensembles de la descendance des femelles accouplées.**

Une classification ascendante hiérarchique des distances euclidiennes et l'analyse en composantes principales répartissent les femelles en 4 sous ensembles

Une fraction importante des femelles mères (N=12, soit 36,40% $\cong$ 2/5) n'engendrera jamais de fils diploïdes. Dans ce sous ensemble, la probabilité de fertilisation de 0,5 conduit à un sex-ratio à la ponte avant toute mortalité de 50:50 c'est à dire autant de fils haploïdes que de filles. Le principe de Fisher (1930), prédisant un investissement égal en fils et filles pour maximiser la transmission des gènes parentaux, intéresserait dans une population à structure d'accouplement mixte (panmictique + local mate competition) seulement 2/5 des femelles mères chez lesquelles aucun croisement entre individus apparentés n'est enviseageable. Ce sous ensemble de femelles est proche de celui théoriquement attendu en présence des croisements entre individus apparentés et non apparentés prévus dans une population où le déterminisme du sexe est du type CSD.

Quand les femelles mères n'engendrent pas de fils diploïdes, la différence observée entre les sex-ratios primaire et secondaire confirme le déficit en fils haploïdes adultes du à un coefficient de survie inférieur à celui des filles. Le reste de la population soit 63,60% de femelles mères ($\cong$ 3/5) se répartit dans trois sous ensembles où des fils diploïdes sont toujours présents : un ovocyte fertilisé sera de sexe mâle avec une probabilité stable de 0,1. Dans ces 3/5 de la population, des accouplements entre parents avec un allèle sexuel en commun sont possibles. La viabilité estimée des ovocytes diploïdes mâle augmente conjointement avec la déviation du sex-ratio en faveur des mâles.. Dans la fraction de la population où la production des fils diploïdes est constante, la déviation progressive du sex-ratio en faveur des mâles (sex-ratio à la ponte ou à l'émergence), est due à la forte baisse des ovocytes fertilisés de sexe femelle parallèle à l'augmentation des ovocytes non fécondés. Ainsi le quatrième sous ensemble composé par 15,10% des femelles se caractérise par une probabilité très faible (0,1) de ponte d'ovocytes fertilisés femelles alors que celle de ponte des ovocytes engagés dans le développement parthénogénètique est une des plus élevées (0,8).

Les paramètres utilisés pour identifier la répartition des femelles accouplées dans 4 sous ensembles montrent l'hétérogénéité de la fonction reproductrice des femelles. Comme l'évolution du sex-ratio du plus de femelles vers le plus de fils suit celle attendue en théorie, on ne peut pas exclure l'hypothèse par laquelle la répartition observée serait due seulement à la structure aléatoire de l'échantillonnage. Dans le cas contraire, il est permis d'envisager des hypothèses impliquant des comportements des femelles fluctuant selon les conditions environnementales (très étudiées chez les parasitoïdes) ou/et une régulation génétique du fonctionnement du complexe de la spermathèque.

## 1. INTRODUCTION

Hymenoptera reproduce by arrhenotoky, males arise from unfertilized eggs and are haploid, whereas females develop from fertilized eggs and are diploid. Three genetic models have been proposed to explain sex determination in hymenoptera: The one-locus multi alleles, the multi-locus multi alleles, and the genic-balance hypotheses [1].

The available data have demonstrated that sex determination is controlled by multi alleles at a single locus in some hymenopteran species. In this one-locus multi alleles model diploid females are considered as being heterozygous for the two alleles wile males can be obtained from unfertilized eggs, (haploid-hemizygous for one sex allele) or from fertilized eggs (diplod-homozygous for one sex allele). [2].

In parasitoids and social Hymenoptera, the presence of diploid males has been studied for the relative frequency of the allele involved in sex determination, the influence they have in the selection of mating structures, sex ratio variability and social behaviour [1, 2, 3, 4, 5]. They are expected to occur at low frequencies in most natural populations of parasitoids and social Hymenoptera and they have been detected in electrophoretic studies [1, 4, 6, 7, 8].

*D. pulchellus* reproduce by arrhenotoky is a solitary endoparasitoid wasp of the lepidopteran *Acrolepiopsis assectella* pupae.

Sex determination in *Diadromus pulchellus*, a solitary endoparasitoid wasp of the lepidopteran *Acrolepiopsis assectella*, follows the CSD model [9]. Experimental crossings with a black-bodied wild strain collected in the south of France and a yellow-bodied strain of *D. pulchellus* carrying the $J^{82}$ mutation have produced 9% of diploid males with an estimated number of 15 alleles in the sample population [9].

Parallel to the work on the sex determination in the hymenopteran *Diadromus pulchellus* [9] we have studied to become it of all the ovocytes laid by unmated and mated females. On the basis of the assumption that in the population used 9% of males can be diploid [9], by viability calculus and theory of probability, we tried to distribute the females mated according to their production of haploid or diploid sons and of diploid daugthers.

Consequently the aim of this work was to determine the fertilisation or non-fertilisation frequencies of ovocytes laid by mated females of *D. pulchellus* and their viability. These estimations include all the eggs laid starting from birth up to the period of peak reproductive activity of females and require a probability model taking into consideration whether an ovocyte had been fertilised or not and its viability irrespective of it being female, diploid male or haploid male. These estimations showed that a mated female during its peak period of reproductive activity functioned on 50% of occasions as a unmated female.

The single locus CSD model of sex determination associated with estimation of fertilisation probability and of viability allowed us to classify the females into 4 groups ; this classification could not have simply resulted from the random nature of the sample.

## 2. MATERIALS AND METHODS
### 2.1. Strains

The hosts were raised in an artificial medium [10], and were all 24h old at the time of parasitism by the females. The parasitoids were raised in a climatic room ; the photoperiod was 16/8h with a day/night temperature of 25 / 15°C, and relative humidity ranging from 50 to 70%. The parasitoid strain was renewed every year with a sample of wild insects collected in the south of France. Seventy females obtained from the second generation of this wild sample were used in the experiments. Thirty-seven of them were maintained as virgins and the 33 others were mated only once at emergence with 3-day-old males.

**2.2. Experimental protocol**

Each host pupa *Acrolepiopsis assectella* allows the development of a single adult parasitoid (the supernumerary parasitoids are eliminated at stage L1 by fights between larvae). In order to prevent superparasitism of the host and to know the fate of each egg during egg-laying, egg-laying was observed under the stereo-microscope. When the ovipositor is introduced into the pupa, the inter-segmental membranes between the sternites 7-8 and 8-9 are extented and permit by transparency the visualisation of the arrival of an ovocyte in the vagina and the propulsion of the ovocyte into the ovipositor. The observation of the egg-laying permits to be sure that a single egg was laid in one host at each sting. The parasitized pupa was then isolated until the emergence of the parasitoid (the sex of the egg was judged only at emergence). Egg-laying was observed during the first 12 days of the life of mated or unmated females because the peak of egg laying is between the 8th and 10th days (with 4 to 5 eggs laid per day) [11]. The egg-laying of each female was observed for 1.5 h every day, and each egg laid was numbered (n°1, 2, 3, 4, etc.). The egg-laying of a female could be shown as an ordered sequence in which the destiny of each egg was listed as follow:

Event A: death of the egg or the L1 larva; (these two states were grouped together);

Event B: death of any one of L2, L3, L4, L5 or of the pre-pupa;

Event C: death of the pupa or the male or female adult inside the host;

Event D: emergence of the male or female adult. Events A, B and C were identified by the dissection of the host when no adult had emerged.

The egg-laying of 70 females were analysed: 37 unmated females laid 1005 unfertilised ovocytes and 33 mated females laid 962 eggs (*table I*).

**Table I**: Egg laying of 37 unmated females and 33 mated females*.

| Unmated females N = 37 | | **Mated females N = 33** | |
|---|---|---|---|
| Events | Eggs | Events | Eggs |
| A | 131 | A | 71 |
| B | 20 | B | 6 |
| C | 8 | C | 20 |
| D | 846 | D males | 405 |
|  |  | D females | 460 |
| Total eggs laid | 1005 | Total eggs laid | 962 |

*Only one egg was laid in each host. Each egg destiny was noted as follows: Event A - Death of the egg or the $L_1$ larva. Event B - Death of the $L_2$ or $L_3$ or $L_4$ or $L_5$ or of the pre-pupa ; Event C - Death of pupa or female or male adult in the host ; Event D - Emergence of male or female adult. Events A, B, and C were identified by dissection of hosts where nothing had emerged.

## 2.3. Models of probability to evaluate the frequency and the viability of the haploid and diploid males

A first set of 3 probabilities ($k_1$, $k_2$, $k_3$) was linked to the sex of the ovocytes laid and a second set ($w_1$, $w_2$, $w_3$) was linked to the viability of the ovocytes laid of which the sex was determined.

• $k_1$ was the probability of an ovocyte being fertilised by a spermatozoon and producing a female egg ;

• $k_2$ was the probability of an ovocyte being fertilised by a spermatozoon and producing a male egg ($k_1$ and $k_2$ were supposed to be different) ;

• $k_3$ was the complementary probability ($k_3 = 1 - k_1 - k_2$) of an ovocyte being laid by a mated female, but remaining unfertilised and developing parthenogenetically into a haploid male.

• $w_1$ was the probability of a diploid female egg producing a viable diploid female adult ;

• $w_2$ was the probability of a diploid male egg producing a viable diploid male adult ;

• $w_3$ was the probability of a haploid male egg producing a viable haploid male adult where the mother could either be mated or virgin. The most accessible probability was $w_3$. This probability was calculated from the time of egg-laying of the virgin females where all the ovocytes laid were unfertilised and had the potential of developing into haploid adult males.

According to the hypothesis of $w_3$ being common to both virgin and mated females.

The progeny of 33 mated adult females was characterised by the list of probabilities $k_1$, $k_2$, $k_3$, and $w_1$, $w_2$, $w_3$. These probabilities were calculated taking into account the future of 962 ovocytes laid by the total number of mated females, and also the number of ovocytes laid by each female.

*2.3.1. Mode of estimation (appendix 1)*

The sum $(k_1 + k_2)$ was estimated first making it possible to deduce $(k_3 = 1 - k_1 - k_2)$. The estimation of $k_1$, $k_2$, $w_1$ and $w_2$, was more difficult. The fact that the probabilities $w_1$, $w_2$ were necessarily superior to 0 and equal or inferior to 1 was used. The first and second hypotheses were based on $w_1=1$ and $w_2=1$, respectively. These two exclusive hypotheses were at the limit of the possible values. The most probable estimation corresponded to the median value, which was equivalent to assuming a symmetric distribution of the estimated values (*table II*).

**Table II**: Median values of fertilisation and viability probabilities of all the ovocytes laid by 33 mated females during the first 12 days of their life *.

|  | $k_1$ | $w_1$ | $k_2$ | $w_2$ | $k_3$ | $w_3$ |
|---|---|---|---|---|---|---|
| hypothesis $w_1 = 1$ | 0.478 | 1 | 0.067 | 0.567 | 0.455 | 0.842 |
| Median values | 0.492 | 0.971 | 0.053 | 0.783 | 0.455 | 0.842 |
| S.E.M. | 0.044 | 0.018 | 0.007 | 0.059 | 0.042 | - |
| hypothesis $w_2 = 1$ | 0.507 | 0.943 | 0.038 | 1 | 0.455 | 0.842 |

* $k_1$ $k_2$ $k_3$ :probabilities linked to the sex of the ovocytes; $w_1$ $w_2$ $w_3 \rightarrow$ probabilities linked to the viability of the ovocytes laid of which the sex was determined ($w_1=1$ or $w_2=1$ were two exclusive hypothese  is at the limit of the possible values ; the most probable estimation corresponds to the median value, which is equivalent to assuming a symmetric distribution of the estimated values). For example: ●) $w_1 = 1$, as $w_3 = 0.842$ (appendix 1) and $k_3 w_3 = 0.383$ (appendix 1) $k_3 = 0.455$ as $k_2 w_2 + k_3 w_3 = 0.421$(appendix 1) $\rightarrow k_2 w_2 = 0.038$ and as $k_1 w_1 = 0.478$ and $k_1 + k_2 = 1- k_3 = 0.545$ (appendix 1) $\rightarrow k_1 = 0.478$, $k_2 = 0.067$, $w_2 = 0.567$ ●) $w_2 = 1$ (using the same reasoning) $\rightarrow w_1 = 0.943$, $k_1 = 0.507$, $k_2 = 0.038$, $k_3 = 0.455$ .

To calculate the median values: $k_1 = (0.507- 0.478) /2 = 0.0145$ and $0.478 + 0.0145 = 0.492$.

**2.4. Distribution of the mated females offspring into four theoretical classes.**

The mating structure of a population plays a major role in the occurrence of diploid males commonly produced by inbreeding [1, 2, 4, 9, 12]. In the single locus CSD system, diploid males were generated from the crossing of parents having a common sex-allele (crossings named matched mating) [13]. It is quite probable that, as for the majority of Hymenoptera, the mating structure of a natural population of the specialist species *D. pulchellus* [14] consists of a combination of panmictic matings (giving unmatched matings) and of local mate competition matings (resulting in matched matings) [5,13]. According to this hypothesis it is possible to associate the theoretical genotypes of the parents with 3 theoretical frequencies of fertilisation of the ovocytes : f = mated ovocytes / total laid ovocytes ; theoretical values are 1, 0 or 1/2, (appendix 2). Under this hypothesis, six families of offspring were expected, each being characterised by the genotypes and the theoretical number of daughters, and haploid and diploid sons (appendix 2). However, as only the phenotype (male or female) and not the genotype was observable, the 6 families were distributed in 4 theoretical classes (I, II, III, IV), each characterised by a sex ratio varying at the one extreme from an exclusive presence of daughters to an exclusive presence of sons at the other extreme, and having a mixture of sons and daughters in between (appendix 2).

According to the hypothesis of a distribution of the offspring of the mated females in 4 theoretical classes, the probabilities of fertilisation and viability of all the ovocytes laid by each of the 33 mated females were analysed using an ascendant and hierarchical classification of the Euclidean distances as well as by an analysis of principal components (PCA). This analysis shows a geometrical representation of the distribution of the offspring of the mated females and the probabilities $w_1$, $w_2$, $k_1$, $k_2$, $k_3$ [15,16]. This made it possible to distribute the 33 mothers in 4 sub-sets characterised by the average value of the 6 probabilities and the primary $[(k_2 + k_3) / k_1]$ and secondary $[(k_2 w_2 + k_3 w_3) / k_1 w_1]$ sex ratios and the sex ratio observed at the birth of the parasitoid adults.

### 3. RESULTS

The survival probability of a haploid male ovocyte was the easiest to calculate. Calculated from the egg-laying of the 37 virgin females, it was related to the ratio of haploid male adults/ the total of haploid ovocytes laid. This probability ($w_3 = 0.842$) corresponded to a constant which was also the survival probability of a male haploid ovocyte laid by a mated female (Appendix 1). In fact, the similar level of egg- laying activity of the two types of females suggested that in the mated females the viability of the haploid male eggs was not affected by the factors of insemination, egg-laying rhythm and the age of the female [11].

## 3.1. Frequency and viability of the diploid and haploid males estimated on the total number of mated females

In a sample of the mated female population, the fertilisation probability of an ovocyte was $k_1 + k_2 = 0.545$ (appendix 1) and the probability of non-fertilisation was $k_3 = 0.455$. With the hypotheses $w_1=1$ and $w_2=1$, and taking into account that for each probability the most probable estimation corresponded to the median value, the estimated values were $k_1 = 0.492$, $k_2 = 0.053$ and $k_3 = 0.455$ (*table II*).

The sex-ratio at egg-laying, [males/females: $(k_2 + k_3) / k_1 = 1.032$] was very close to 1 with $k_2 + k_3 = 0.508$ and $k_1 = 0.492$ (*table II*). The survival probability of a female egg was 0.971, of a diploid male egg 0.783, and it was 0.842 for an unfertilised ovocyte (*table II*). Although the lowest value was that of the diploid males, the survival probabilities were high for the 3 types of 'eggs'. In these conditions, the calculated sex ratio at emergence was 0.90: $[[(k_2 w_2) + (k_3 w_3)] / (k_1 w_1) = 0.4246 / 0.4777 = 0.8889 \cong 0.90]$. The slight difference between the calculated sex ratio at egg-laying and at emergence, showed a low mortality in the period between these two events (*table II*).

## 3.2. Estimations for each of the mated females: identification of 4 sub-sets of mated females

Under the hypothesis that the offspring of mated females can be classified in 4 theoretical groups, the probabilities of fertilisation and of viability of all the ovocytes laid by each of the 33 mated females ($w_1$, $w_2$, $k_1$, $k_2$, $k_3$.) (*table III*) were analysed by an ascending hierarchical classification of Euclidean distances and by an analysis of their principal components (*figure 1* and *table IV*) [15,16]. The offspring of the 33 mothers were distributed in 4 sub-sets, and each set was characterised by the average values of the probabilities of fertilisation, viability and by the primary $[(k_2 + k_3) / k_1]$ and the secondary $[(k_2 w_2 + k_3 w_3) / k_1 w_1]$ calculated sex ratios and the observed sex-ratio (*table IV*).

**Table III:** Probabilities of fertilisation and of viability ($w_1$, $w_2$, $k_1$, $k_2$, $k_3$) of all the eggs laid by each of the 33 mated females

| Numbers of mated females | $w_1$ | $w_2$ | $k_1$ | $k_2$ | $k_3$ |
|---|---|---|---|---|---|
| 1 | 0.88 | 0.28 | 0.57 | 0.10 | 0.33 |
| 2 | 0.89 | 0.10 | 0.78 | 0.10 | 0.12 |
| 3 | 1.00 | 1.00 | 0.74 | 0.01 | 0.25 |
| 4 | 0.97 | 0.78 | 0.45 | 0.05 | 0.50 |
| 5 | 0.84 | 0.31 | 0.54 | 0.10 | 0.36 |
| 6 | 0.94 | 0.12 | 0.85 | 0.06 | 0.09 |
| 7 | 0.96 | 0.27 | 0.81 | 0.05 | 0.14 |
| 8 | 1.00 | 1.00 | 0.66 | 0.00 | 0.34 |
| 9 | 0.66 | 0.83 | 0.04 | 0.09 | 0.87 |
| 10 | 1.00 | 1.00 | 0.10 | 0.09 | 0.81 |
| 11 | 0.98 | 0.61 | 0.69 | 0.04 | 0.27 |
| 12 | 0.95 | 0.61 | 0.86 | 0.06 | 0.08 |
| 13 | 1.00 | 1.00 | 0.44 | 0.02 | 0.54 |
| 14 | 1.00 | 1.00 | 0.27 | 0.09 | 0.64 |
| 15 | 0.90 | 0.18 | 0.73 | 0.09 | 0.18 |
| 16 | 0.91 | 0.64 | 0.33 | 0.08 | 0.59 |
| 17 | 1.00 | 1.00 | 0.55 | 0.03 | 0.42 |
| 18 | 1.00 | 1.00 | 0.71 | 0.02 | 0.27 |
| 19 | 1.00 | 1.00 | 0.19 | 0.09 | 0.72 |
| 20 | 1.00 | 1.00 | 0.31 | 0.02 | 0.67 |
| 21 | 1.00 | 1.00 | 0.29 | 0.09 | 0.62 |
| 22 | 1.00 | 1.00 | 0.50 | 0.00 | 0.50 |
| 23 | 0.96 | 0.18 | 0.87 | 0.04 | 0.09 |
| 24 | 1.00 | 1.00 | 0.44 | 0.09 | 0.47 |
| 25 | 1.00 | 1.00 | 0.57 | 0.00 | 0.43 |
| 26 | 0.99 | 0.53 | 0.84 | 0.02 | 0.14 |
| 27 | 1.00 | 1.00 | 0.04 | 0.09 | 0.87 |
| 28 | 0.76 | 0.39 | 0.31 | 0.12 | 0.57 |
| 29 | 0.91 | 0.35 | 0.59 | 0.08 | 0.33 |
| 30 | 0.71 | 0.18 | 0.47 | 0.17 | 0.36 |
| 31 | 1.00 | 1.00 | 0.27 | 0.03 | 0.70 |
| 32 | 0.65 | 0.74 | 0.07 | 0.09 | 0.84 |
| 33 | 0.90 | 0.53 | 0.39 | 0.08 | 0.53 |

The distribution of the offspring of the 33 mothers in 4 sub-sets was heterogeneous since 12 individuals were each placed in sub-sets n°1 and n°2 and only four and five females were in sets n°3 and n°4, respectively (*table IV*). The most representative ones were those where the sex ratio was very much in favour of the daughters (n°1) or led to equality (n°2) between haploid sons and daughters (*table IV*).

**Table IV:** Under the hypothesis of the offspring of mated females being classified in 4 theoretical classes I, II, II, IV (appendix 2)*.

**a.**

| Sub sets | numbers of mothers | Identity of mated females in each sub-set |
|---|---|---|
| 1 | 12 | 1 -2 - 5 - 6 - 7 - 11 - 12 -15 - 23 - 26 - 29 - 30 |
| 2 | 12 | 3 - 8 - 13 - 14 - 17 - 18 - 20 - 21 - 22 - 24 -25 - 31 |
| 3 | 4 | 4 - 16 - 28 - 33 |
| 4 | 5 | 9 - 10 - 19 - 27 - 32 |

**b.**

| Sub sets | numbers of mothers | $k_1$ | $w_1$ | $k_2$ | $w_2$ | $k_3$ | $w_3$ | sex-ratios primary calculated 1 | secondary calculated 2 | observed 3 |
|---|---|---|---|---|---|---|---|---|---|---|
| 1 | 12 | 0.7 | 0.9 | 0.1 | 0.3 | 0.2 | 0.842 | 0.428 | 0.310 | 0.342 |
| 2 | 12 | 0.5 | 1.0 | 0.0 | 1.0 | 0.5 | 0.842 | 1 | 0.842 | 0.849 |
| 3 | 4 | 0.4 | 0.9 | 0.1 | 0.6 | 0.5 | 0.842 | 1.5 | 1.33 6 | 1.418 |
| 4 | 5 | 0.1 | 0.9 | 0.1 | 0.9 | 0.8 | 0.842 | 9 | 8.484 | 5.363 |

- The ascendant and hierarchical classification of the Euclidean distances as well as the analysis of principal components distributed the females in 4 sub-sets (1, 2, 3, 4). (PCA : Classification on lines, Ascending hierarchical classification, criterion of aggregation : average of the balanced distances) . **a** - Identity of mated females in each sub-set (figure 1); **b -** Each set was characterised by the average values of the frequencies of fertilisation and viabilities and by the sex-ratios. The primary calculated sex-ratio (1) ($k_2+k_3/k_1$), the secondary (2) calculated sex ratio using the fertilisation and viability frequencies ($k_2w_2+ k_3w_3/k_1w_1$), and the observed sex ratio (3) known at the emergence of the parasitoid adults.

The females (2/5 of them) which would never produce any diploid sons were placed in the latter category. This set was close to the theoretical class II (family 3, appendix 2). In this $3^r$ theoretical family, the probability of fertilisation of 0.5 led to a sex ratio of 50/50 at egg laying, i.e, equal numbers of daughters and haploid sons.

Some ovocytes which would develop parthenogenetically were expected to be in the $4^{th}$ sub-set according to a probability $k_3$ varying from 0.2 to 0.8 (*table IV*). The increase in the probability of non-fertilisation was parallel to the decrease in the probability of fertilisation of an ovocyte. This led to an evolution of the calculated and observed sex ratio favouring males (*table IV*). This evolution went from a nearly exclusive presence of female offspring to an exclusive presence of haploid male offspring. The projection on a reference plane defined by x-axis (females/males) and y-axis (haploidy-diploidy) clearly showed the grouping of the mated females going from more daughters towards less daughters (*figure 1*). This development was close to the one anticipated by the theoretical grouping of the offspring expected in a mixed structure, grouping matched and unmatched matings

**Figure 1.** Under the hypothesis of simple locus determination model and the offspring of mated females being classified in 4 theoretical sub-sets (sub-sets 1, 2, 3, 4.) (see, Table 2).

Distribution of the offspring of the 33 mated females resulting from an analysis of the principal components (numbers 1, 2, 3,...33 correspond to the identity of each tested female).

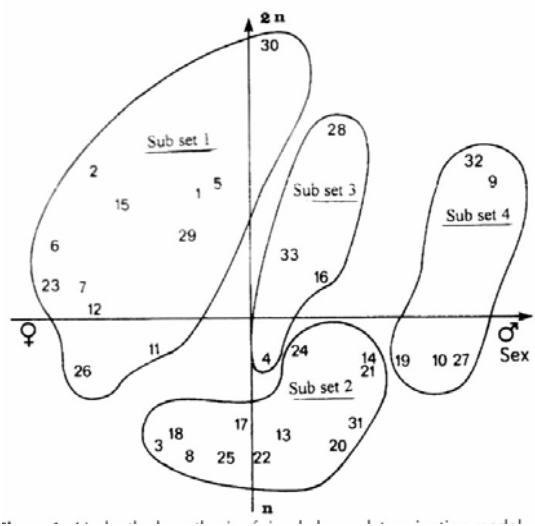

## 4. DISCUSSION

From the total number of *D. pulchellus* mated females, the estimation of the fertilisation parameter [probability of fertilisation of the female and male ovocytes ($k_1 + k_2 = 0.545$)] demonstrated that half of the ovocytes laid were fertilised. The calculated values of each parameter revealed that the probability of fertilisation of a male ovocyte ($k_2 = 0.053$) is 10 times less than that of a female ($k_1 = 0.492$). To this weak proportion of diploid male ovocytes was added the number of unfertilised ovocytes ($k_3 = 0.455$) ready for a parthenogenetic development. Estimated from the period of maximum egg-laying activity, these probabilities showed that in *D. pulchellus* mated females functioned 50% of the time as virgin females (female ovocytes $k_1 = 0.492$, haploid male ovocytes $k_3 = 0.456$). This revealed that a mated female could regulate its stock of spermatozoa accumulated after mating. A quantitative study of this regulation by the eupelmid *Eupelmus orientalis*, shows that the progressive use of spermatozoa is a function of egg-laying activity and the age of the mother [17]. In mated females of *D. pulchellus*, this regulation of spermatozoa leads to the production of daughters at the end of the life of a female [11]. However until now, haploid males were supposed to be the exclusive offspring of virgin and mated females, the latter having partly exhausted their stock of spermatozoa [18, 19].

At egg-laying the estimated sex ratio in the population sample was close to 1 [sons (2n +n) / daughters = 1.032] and showed a phenotypic equality between the sexes. This concealed an unequal reproductive cost between the two categories of sons because the production of diploid sons requires the mobilisation of the spermatozoa and of all the secretions favouring the fertilisation of ovocytes [20, 21].

The egg-laying of one ovocyte per host excludes mortality due to competition among neonatal larvae, but the development of an egg to the adult stage can be disturbed at the embryonic, larval or pupal stages. In the absence of superparasitism a laid ovocyte, whatever its sex might be, had a high probability of developing to the adult state because the survival probability of a female egg was 0.971, of a diploid male egg was 0.783 and of an unfertilised egg was 0.842. The calculated sex ratio of 0.90 at emergence showed a low mortality level since the survival probability of the ovocytes laid was taken into account in its calculation. Estimated from the total number of offspring of all mated females, this calculated sex ratio at emergence compared to that at egg- laying  (= 1), revealed a low deficit in haploid and diploid male adults. This confirms the observation that in *D. pulchellus,* even in the absence of larval competition, mortality mainly affects the males [22].

The estimation of the probability of fertilisation and the viability of eggs laid by each of the 33 mated females allowed the identification of 4 sub-sets of females. A large portion of the mothers (36.40 % $\cong$ 2/5) do not produce diploid sons. In this sub-set the fertilisation probability (0.5) led to a 50:50 sex ratio at egg-laying before mortality, thus producing a similar number of haploid sons and daughters. The Fisher principle [23], predicting an equal investment in daughters and sons to maximise the transmission of the parental genes, will affect a population (with a mixed mating structure: panmictic + local mate competition) ; only unmatched mating was possible for these mothers.

When mothers did not give birth to diploid sons, the observed difference between the calculated primary and secondary sex-ratios confirmed the deficit in haploid adult sons due to a survival coefficient inferior to that of daughters. In the absence of competition, the differential mortality can be explained; since the males are haploid, all lethal mutations are expressed in the phenotype [22].

The rest of the mothers (63.60% = 3/5) were distributed in the remaining 3 sub-sets where diploid sons were always present; one fertilised ovocyte will be male with a stable probability of 0.1. In 3/5 of the population, matings between parents having a common sex allele were possible. The estimated viability of the diploid male ovocytes increased together with the sex-ratio deviation towards the males. These diploid males mainly produce diploid sperm which can penetrate the ovocyte [24], but the absence of daughters associated with a high level of mortality suggests that triploid females normally die during the early stages of development [25]. This observation corroborates other studies made in the Hymenoptera [1,2], and the idea that most of the *D. pulchellus* diploid males are sterile. However, in *D. pulchellus,* diploid males can also exceptionally produce haploid [25]. This haploid sperm on penetrating the ovocyte, gives rise to diploid females which represent $1°/_{00}$ of the offspring (8 diploid daughters / 939) [25].

In the part of the population where the production of diploid males was constant, the progressive deviation of the sex-ratio favouring males (sex-ratio at egg-laying or at emergence) was due to a considerable decrease in the fertilised female ovocytes in parallel with an increase in the unfertilised ovocytes. Thus, the sub-set consisting of 15.10% of females, was characterised by a very small fertilised female egg-laying probability (0.1), although the probability of the laying of ovocytes engaged in parthenogenetic development was one of the highest (0.8).

The parameters used to identify the distribution of mated females in 4 sub-sets showed the heterogeneity of the female reproductive function. As the sex ratio evolved from a predominantly female to a predominantly male one (as expected theoretically), one can hypothesise that the classification observed could not have simply resulted from the random nature of the sample. In the opposite situation, it was possible to put forward some hypotheses involving female behaviour fluctuating with environmental conditions, a phenomenon which has been widely studied in parasitoids [4,26,27], and/or a genetic regulation of the functioning of the spermatheca complex. This organ has a complex morphological and anatomical structure [20, 21] which may vary in the coordination of its function in a population of females. This variability could be the expression of a genetic regulation of the spermatheca-spermatozoa complex, which could be difficult to demonstrate experimentally.

**Acknowledgments -** The authors would like to thank Dr. Michel Gillois for his helpful suggestion on the models of probabilities, and Professor Vincent Labeyrie for stimulating discussion.

**Appendix 1 : Estimation of the fertilisation and viability probabilities**

Estimation of $w_3$ (values : Table 1))

$w_1$ was the probability of a diploid female egg giving a viable diploid female adult.

$w_2$ was the probability of a diploid male egg giving a viable diploid male adult.

$w_3$ was the probability of a haploid male egg giving a viable haploid male adult where the mother could either be mated or virgin.

$$w_3 = \frac{\text{total haploid adult males developing up to event D}}{\text{total haploid ovocytes layed by 37 unmated females}}$$

$$w_3 = \frac{846}{1005} = 0{,}842$$

Estimation of $k_1$, $k_2$, $k_3$, $w_1$, $w_2$ (values Table 1)

N $\varpi$ : total ovocytes laid (fertilised and unfertilised) = 962

    $\rightarrow$ *composed of:*

N1 $\varpi$ : fertilised ovocytes (2n) producing a female egg.

N2 $\varpi$ : fertilised ovocytes (2n) producing a male egg.

N3 $\varpi$ : unfertilised ovocytes (n) producing a male egg.

→ *with the following relations:*

N $\varpi$     gives $N_{ad}$ (ad = adults)

N1 $\varpi$     gives $N_1 ad$ females 2n

N2 $\varpi$     gives $N_2 ad$ males 2n

N3 $\varpi$     gives $N_3 ad$ males n

→ *knowing that:*

N $\varpi$ = total ovocytes laid = 962 and total number of adults = 460 + 405 = 865

(1) $N \varpi \cdot [k_1 w_1] = N_1 ad = 460$

(2) $N \varpi \cdot [k_2 w_2] + N \varpi [1-k_1-k_2]w_3 = N_2 ad + N_3 ad = 405$

(3) $N \varpi [k_1 w_1 + k_2 w_2] + N \varpi [1-k_1-k_2]w_3 = N_1 ad + N_2 ad + N_3 ad = N_{ad} = 865$

(4) $N \varpi [k_1 w_1 + k_2 w_2 + (1 - k_1 - k_2) w_3] = 865$

(5) $\dfrac{N_{ad}}{N \varpi} = \dfrac{865}{962} = k_1 w_1 + k_2 w_2 + (1 - k_1 - k_2) w_3$

with $w_3 = 0{,}842$    $\dfrac{N_{ad}}{N \varpi} = k_1 w_1 + k_2 w_2 + k_3 w_3$

    $k_3 = 1 - k_1 - k_2$

→ *by using*

(6) $\dfrac{N_{ad}}{N \varpi} = 0{,}899$

    $\dfrac{N_{1,ad}}{N \varpi} = k_1 w_1 = \dfrac{460}{962} = 0{,}478$

(7) $\dfrac{N_{2,ad} + N_{3,ad}}{N \varpi} = k_2 w_2 + k_3 w_3 = \dfrac{405}{962} = 0{,}421$

→ proportion of adult males = (2n + n) in the population of *Diadromus pulchellus*

$$\left(\dfrac{\text{males}}{\text{males + females}}\right) = \dfrac{0{,}421}{0{,}478 + 0{,}421} = \dfrac{0{,}421}{0{,}899} = 0{,}468$$

→ for Periquet et al. (1993), the percentage of males 2n in the population of *Diadromus pulchellus* was 9%.

$$\dfrac{N_{2,ad}}{N_{2,ad} + N_{3,ad}} = 0{,}09 = \dfrac{k_2 w_2 \cdot N \varpi}{(k_2 w_2 + k_3 w_3) N \varpi} = \dfrac{k_2 w_2}{k_2 w_2 + k_3 w_3}$$

$$k_2 w_2 = 0{,}09 \, [k_2 w_2 + k_3 w_3]$$

$k_2 w_2 [1 - 0,09] = k_3 w_3 \ 0,09$

$0,91 \ k_2 w_2 = k_3 w_3 \ 0,09$

using the equation n° 7   $k_2 w_2 + k_3 w_3 = 0,421$

thus   $k_2 w_2 + k_3 w_3 = 0,421$ $\left.\begin{array}{l}\\ \\ \\ \end{array}\right\} \Rightarrow k_3 w_3 = 0,383$ as $w_3 = 0,842$

$\quad 0,91 \ k_2 w_2 - k_3 w_3 \ 0,09 = 0$

$\Rightarrow k_3 = 0,455$

$\Rightarrow k_2 w_2 = 0,038$

and $\dfrac{N_{1,ad}}{N\varpi} = k_1 w_1 = \dfrac{460}{962} = 0,478$

**Appendix 2 : Unmatched and matched matings under single locus (CSD)**

| | Unmatched matings | | | | matched matings | | |
|---|---|---|---|---|---|---|---|
| Crosses | mother $X_i X_j$ | father $X_k$ | | | mother $X_i X_j$ | | father $X_i$ |
| Offspring | daughters 2n $X_i X_k$ | sons n $X_i$ or $X_j$ | | | daughters 2n $X_i X_j$ | sons 2n $X_i X_i$ | sons n $X_i$ or $X_j$ |
| Theoretical frequencies of fertilisation | Expected offspring | | | | | | |
| | daughters 2n | sons n | Families | daughters 2n | sons 2n | sons n | |
| f = 1 | 100 | 0 | → 1 - 2 ← | 50 | 50 | 0 | |
| f = 0.5 | 50 | 50 | → 3 - 4 ← | 25 | 25 | 50 | |
| f = 0 | 0 | 100 | → 5 - 6 ← | 0 | 0 | 100 | |

a) In this model, sex determination results from the complementary action of two different alleles : diploid females are heterozygous for two alleles ($X_i X_j$) while haploid and diploid males are hemizygotes ($X_i$ or $X_j$) and homozygotes ($X_i X_i$ or $X_j X_j$) respectively. Six families of offspring were expected from the association of the theoretical genotypes of the parents with the 3 theoretical frequencies of fertilisation.

| Groups of families | Theoretical classes | Frequencies of fertilisation | | Theoretical primary sex-ratio sons (2n+n) / daughters 2n |
|---|---|---|---|---|
| 1 | I | daughters 2n | f = 1 | exclusively daughters 2n |
| 2 + 3 | II | daughters 2n, sons 2n, sons n | $f = 0.5$<br>$f = 1$ | 50 : 50 |
| 4 | III | daughters 2n, sons 2n, sons n | $f = 0.5$ | 75 : 25 |
| 5 + 6 | IV | sons n | $f = 0$ | exclusively sons n |

b) As only the phenotype (male or female) and not the genotype was observable, the 6 families were distributed in 4 theoretical classes (I - II - III - IV).